\documentclass[pra,showpacs,preprintnumbers,amsmath,aps]{revtex4}
\usepackage{graphics}

\def\be{\begin{equation}}
\def\ee{\end{equation}}
\def\bea{\begin{eqnarray}}
\def\eea{\end{eqnarray}}
\def\bse{\begin{subequations}}
\def\ese{\end{subequations}}
\def\bma{\begin{mathletters}}
\def\ema{\end{mathletters}}
\def\C{\hbox{$\mit I$\kern-.6em$\mit C$}}

\begin{document}

\title{Mixture of multiple copies of maximally entangled states is quasi-pure}
\author{Dong Yang}
\email{dyang@zju.edu.cn} 
\author{Yi-Xin Chen}
\affiliation{Zhejiang Institute of Modern Physics and
Department of Physics, Zhejiang University, Hangzhou 310027, P. R. China}

\date{\today}

\begin{abstract}
Employing the general BXOR operation and local state discrimination, the mixed state of the form 
\be
\rho^{(k)}_{d}=\frac{1}{d^{2}}\sum_{m,n=0}^{d-1}(|\phi_{mn}\rangle\langle\phi_{mn}|)^{\otimes k} \nonumber
\ee
is proved to be quasi-pure, where $\{|\phi_{mn}\rangle\}$ is the canonical set of mutually orthogonal maximally entangled states in $d\times d$. Therefore irreversibility does not occur in the process of distillation for this family of states. Also, the distillable entanglement is calculated explicitly. 
\end{abstract}


\maketitle

Entanglement distillation plays a crucial role in quantum information theory \cite{Bennett1}. As well-known, entanglement is responsible for many quantum tasks and pure entangled states are required in most cases. Unfortunately, due to the interaction with the environment, pure entanglement is fragile and easy to be blurred by the noise. So distillation of entanglement is of importance. Though many distillation protocols and upper bounds are known, distillable entanglement are known in few nontrivial cases \cite{Rains}. Actually, entanglement distillation is closely related to local state discrimination. Two counter-intuitive facts are found about local state discrimination. One is that there exist product orthogonal states that could not be discriminated exactly by LOCC operations \cite{Bennett}. The other is that any two orthogonal multipartite states could be discriminated with certainty by only LOCC operations \cite{Walgate1}. Based on the results of \cite{Walgate1}, the distillable entanglement \cite{Bennett1} is calculated for a class of mixed states comprising of Bell basis \cite{Ghosh,Yang}. For generic bipartite mixed states, entanglement distillation is an irreversible process in the sense that more pure-state entanglement is required to create it than can be distilled from it in the asymptotic limit \cite{Vidal}. However, the process of distillation is reversible in the so-called quasi-pure mixed state \cite{Vollbrecht}. In this paper, the mixture of multiple copies of maximally entangled states with equal probability \cite{Hamieh} is proved to be quasi-pure. So the distillation process is reversible that is not so obvious at first glance. Furthermore, the distillable entanglement is calculated explicitly.

We consider the mixed state of the form,
\be
\rho^{(k)}_{d}=\frac{1}{d^{2}}\sum_{m,n=0}^{d-1}(|\phi_{mn}\rangle\langle\phi_{mn}|)^{\otimes k},
\ee 
where $\{|\phi_{mn}\rangle\}$ is the canonical set of mutually orthogonal maximally entangled states(MES) in $d\times d$ defined as,  
\be
|\phi_{mn}\rangle=\frac{1}{\sqrt{d}}\sum_{j=0}^{d-1}e^{2\pi ijn/d}|j\rangle|j\oplus m\rangle, n,m=0,1,\cdots,d-1.
\ee
where $\oplus$ means addition modulo $d$.

The situation occurs where Alice and Bob share $k$ pairs of qudits, each pair in the same maximally entangled state. In order to share entanglement, they should store the classical information in classical memory that identifies in which state the $k$ pairs of qudits are. Now imagine that by some misfortune, the data in the classical memories are deleted, how many ebits can be distilled? As well-known, it is difficult to calculate the distillable entanglement of the mixed states. The main result of the paper is that the mixed state of $Eq.(1)$ is quasi-pure \cite{Vollbrecht}. For quasi-pure states, irreversibility does not occur in the entanglement distillation. And the distillable entanglement is easy to calculated. Before the proof of the main result, let us explain what is the meaning of "quasi-pure".

A mixed state is quasi-pure if it can be reversibly transformed to the following form under the local operation and classical communication (LOCC),
\be
\rho=\sum p_{i}\rho_{i}\otimes|\phi_{i}\rangle\langle\phi_{i}|,
\ee 
where $\rho_{i}$ are separable and orthogonal states that can be discriminated exactly by LOCC operations. The physical meaning of the quasi-pure state is that every pure entangled state is identified by a separable "tag state" that can be distinguished by LOCC operations. As the transformation is reversible, all the entanglement in the original mixed state is contained in the pure states and can be distilled. As Bennett et. al \cite{Bennett} showed that even separable orthogonal states might contain nonlocal property that can not be identified by LOCC operations, it is necessary to introduce the "tag state" distinguishable by LOCC operations. Similarly, we can define separable-quasi-pure and PPT-quasi-pure with respect to the operations allowed. For instance, consider the constructed state $\rho=1/9\sum |\psi_{i}\rangle\langle\psi_{i}|\otimes|\phi_{i}\rangle\langle\phi_{i}|$,   
where $|\psi_{i}\rangle$ is the separable pure orthogonal basis in $3\times 3$ in \cite{Bennett} and $|\phi_{i}\rangle$ is orthogonal MES in $3\times 3$. Explicitly, this state is separable-quasi-pure. And the distillation is a reversible process if separable operations are allowed. Is it also LOCC-quasi-pure? Does irreversibility occur in this mixed state? If it does, we can conclude that the irreversibility originates from the nonlocal property of the separable basis. Though we conjecture it is the case, the strict proof escape from us. However, we will show that the state of $Eq.(1)$ is quasi-pure though it looks unlike.

{\bf Main Result}: The mixed state of maximally entangled states in $Eq.(1)$ is quasi-pure. 

First we introduce the operations that are employed. 
The control-not operation $C$ is defined as
$C|i\rangle\otimes|j\rangle=|i\rangle\otimes|j\oplus i\rangle$, and the bilateral control-not operation (BXOR) $B$ is 
$B|i\rangle_{A1}|r\rangle_{B1}\otimes|j\rangle_{A2}|s\rangle_{B2}=|i\rangle_{A1}|r\rangle_{B1}\otimes|j\oplus i\rangle_{A2}|s\oplus r\rangle_{B2}$,
where $\{|i\rangle\}$ is the computational basis set. Denote $B(m,n)$ as the BXOR operation performed on the mth pair (source) and the nth pair (target), and $\mathcal{B}$ act on $k$ copies of $\phi_{mn}$ as 
\bea
\mathcal{B}\phi_{mn}^{\otimes k}=B(1,k)B(2,k)\cdots B(k-1,k)\phi_{mn}^{\otimes k}=\phi_{m,0}^{\otimes (k-1)}\phi_{\oplus km,n}.
\eea  

For brevity, we denote the projector $|\phi_{mn}\rangle\langle\phi_{mn}|$ as $P_{mn}$ and omit the normalization for the state unless confusion occurs. Three important properties about $|\phi_{mn}\rangle$ are utilized. First, $\sum_{m}P_{mn}$ and $\sum_{n}P_{mn}$ are separable. For example, the subspace spanned by $\{\phi_{0n}\}$ can also be spanned by $\{|ii\rangle, i=0,\cdots,d-1\}$. For the subspace spanned by $\{\phi_{m0}\}$, it is not so obvious, but can be proved to be separable in the dual basis $\{|e_{k}^{A}\rangle\}$, $\{|e_{k}^{B}\rangle\}$, where $|k^{A}\rangle=1/\sqrt{d}\sum_{j=0}^{d-1}e^{2\pi ijk/d}|e_{j}^{A}\rangle$ and $|e_{k}^{B}\rangle=1/\sqrt{d}\sum_{j=0}^{d-1}e^{2\pi ijk/d}|j^{B}\rangle$. Second, $\sum_{m}P_{mn}$ are orthogonal to each other and can be distinguished exactly by LOCC. So do $\sum_{n}P_{mn}$. Third, for fixed $n$ and varying $m$, $|\phi_{mn}\rangle$ can be transformed to $|\phi_{0m}\rangle$ by local basis transformation.

Proof:
\be
\rho^{(k)}_{d}\stackrel{\mathcal{B}}{\longrightarrow}\sum_{m}P_{m0}^{\otimes (k-1)}\sum_{n}P_{\oplus km,n}.
\ee 
Denote the maximal common factor between $k$ and $d$ as $g(k,d)$. When $g(k,d)=1$,
We know that $\{\oplus km, m is varying\}$ is a permutation of the numbers $\{0,1,\cdots,d-1\}$. The $d$ mixed states $\sum_{n}P_{\oplus km,n}$ are separable and orthogonal to each other. So $\rho^{(k)}_{d}$ is quasi-pure. When $g(k,d)>1$, the problem is a little complicated. Suppose $k=g*\tilde{k}, d=g*\tilde{d}$, where $\tilde{k}, \tilde{d}$ are integers and satisfy $g(\tilde{k},\tilde{d})=1$. For $m=0,1,\cdots,d-1$, it can be written as $m=s\tilde{d}+t, s=0,1,\cdots,g-1, t=0,1,\cdots,\tilde{d}-1$. Then
\be
\oplus_{d}(km)=\oplus_{(g\tilde{d})}g\tilde{k}(s\tilde{d}+t)=\oplus_{(g\tilde{d})}g\tilde{k}t=g\oplus_{\tilde{d}}\tilde{k}t.
\ee
Because $g(\tilde{k},\tilde{d})=1$, $\{\oplus_{\tilde{d}}\tilde{k}t, t=0,1,\cdots,\tilde{d}-1\}$ is a permutation of the numbers $\{0,1,\cdots,\tilde{d}-1\}$. For brevity, set $T(t)=g\oplus_{\tilde{d}}\tilde{k}t$ that is different for $t=0,1,\cdots,\tilde{d}-1$.
So
\be
\sum_{m}P_{m0}^{\otimes (k-1)}\sum_{n}P_{\oplus km,n}=\sum_{t}\{\sum_{s}P_{s\tilde{d}+t,0}^{\otimes k-1}\}\{\sum_{n}P_{T(t),n}\}.
\ee
For every $t$, $\sum_{n}P_{T(t),n}$ is separable and orthogonal and can be discriminated by local operations. Now, we show that $\sum_{s}P_{s\tilde{d},0}^{\otimes (k-1)}$ is quasi-pure since for the other cases $t\neq 0$, $\sum_{s}P_{s*\tilde{d}+t,0}^{\otimes (k-1)}$ can be transformed to $\sum_{s}P_{s*\tilde{d},0}^{\otimes (k-1)}$ by the permutation of the basis states. 
\be
P_{s*\tilde{d},0}(d)=P_{s,0}(g)P_{00}(\tilde{d}),
\ee
where $P_{s,0}(g)$ represents the projector of the MES in $g\times g$.
So
\be
\sum_{s}P_{s*\tilde{d},0}^{\otimes (k-1)}=\sum_{s}P_{s,0}^{\otimes (k-1)}(g)P_{00}^{\otimes (k-1)}(\tilde{d}).
\ee
Under the dual basis, $\sum_{s}P_{s,0}^{\otimes (k-1)}(g)$ can be written as $\sum_{s}P_{0,s}^{\otimes (k-1)}(g)$. Further employing the operation $\mathcal{B}$
\be
\sum_{s}P_{0,s}^{\otimes (k-1)}(g)\stackrel{\mathcal{B}}{\longrightarrow}P_{0,s}^{\otimes (k-2)}(g)\sum_{s}P_{0,s}(g)
\ee
So $\sum_{s}P_{s*\tilde{d},0}^{\otimes (k-1)}$ is quasi-pure. As all the operations are local reversible operations, the original state is quasi-pure. Though measurements are performed during the process, they are non-demolation measurements and do not destroy the state.

In the following, we give some examples and calculate the distillable entanglement $E$ that is equal to the entanglement cost as the mixed state is quasi-pure.

{\bf Ex1} $d=2$
\begin{table}[htbp]
\begin{tabular}{|c|c|c|c|c|c|}\hline
$k$&$\phi_{00}^{\otimes k}$&$\phi_{01}^{\otimes k}$&$\phi_{10}^{\otimes k}$&$\phi_{11}^{\otimes k}$&$E(\rho_{d}^{(k)})$
\\ \hline\hline
$1$&$\phi_{00}$&$\phi_{01}$&$\phi_{10}$&$\phi_{11}$&$0$
\\ \hline
$2$&$\phi_{00}\otimes\phi_{00}$&$\phi_{00}\otimes\phi_{01}$&$\phi_{10}\otimes\phi_{00}$&$\phi_{10}\otimes\phi_{01}$&$0$
\\ \hline
$3$&$\phi_{00}^{\otimes 2}\otimes\phi_{00}$&$\phi_{00}^{\otimes 2}\otimes\phi_{01}$&$\phi_{10}^{\otimes 2}\otimes\phi_{10}$&$\phi_{10}^{\otimes 2}\otimes\phi_{11}$&$2\log2$
\\ \hline
$4$&$\phi_{00}^{\otimes 3}\otimes\phi_{00}$&$\phi_{00}^{\otimes 3}\otimes\phi_{01}$&$\phi_{10}^{\otimes 3}\otimes\phi_{00}$&$\phi_{10}^{\otimes 3}\otimes\phi_{01}$&$2\log2$
\\ \hline
\end{tabular}
\caption{Under the operation $\mathcal{B}$, what happens to $k$ copies of MES in $2\times 2$. The last column show the entanglement of $\rho^{(k)}_{2}$.}
\end{table}

\bse
\bea
&\rho^{(2)}_{2}&\stackrel{\mathcal{B}}{\longrightarrow}(P_{00}+P_{10})(P_{00}+P_{01}),\\
&\rho^{(3)}_{2}&\stackrel{\mathcal{B}}{\longrightarrow}P_{00}^{\otimes 2}(P_{00}+P_{01})+P_{10}^{\otimes 2}(P_{10}+P_{11}),\\
&\rho^{(4)}_{2}&\stackrel{\mathcal{B}}{\longrightarrow}(P_{00}^{\otimes 3}+P_{10}^{\otimes 3})(P_{00}+P_{01})
\eea
\ese

Because $P_{00}+P_{10}$, $P_{00}+P_{01}$ are separable, $\rho^{(2)}_{2}$ is separable. The same conclusion is obtained through direct calculation \cite{Smolin}. $E(\rho^{(2)}_{2})=0$.
Similarly it is easy to show $\rho^{(3)}_{2}$ is quasi-pure and $E(\rho^{(3)}_{2})=2$. 
It is required to explain for $\rho^{(4)}_{2}$. $P_{00}^{\otimes 3}+P_{10}^{\otimes 3}$ can be transformed to $P_{00}^{\otimes 3}+P_{01}^{\otimes 3}$ by local unitary operation. Performing $\mathcal{B}$ again,
$P_{00}^{\otimes 3}+P_{01}^{\otimes 3}\stackrel{\mathcal{B}}{\longrightarrow}P_{00}^{\otimes 2}(P_{00}+P_{01})$
that is quasi-pure and $E(\rho^{(4)}_{2})=2$. For larger $k$, the cases above occur similarly.

{\bf Ex2} $d=3$
\begin{table}[htbp]
\begin{tabular}{|c|c|c|c|c|}\hline
$k$&$\phi_{0n}^{\otimes k}$&$\phi_{1n}^{\otimes k}$&$\phi_{2n}^{\otimes k}$&$E(\rho_{d}^{k})$ 
\\ \hline\hline
$1$&$\phi_{0n}$&$\phi_{1n}$&$\phi_{2n}$&$0$
\\ \hline
$2$&$\phi_{00}\otimes\phi_{0n}$&$\phi_{10}\otimes\phi_{2n}$&$\phi_{20}\otimes\phi_{1n}$&$\log3$
\\ \hline
$3$&$\phi_{00}^{\otimes 2}\otimes\phi_{0n}$&$\phi_{10}^{\otimes 2}\otimes\phi_{0n}$&$\phi_{20}^{\otimes 2}\otimes\phi_{0n}$&$\log3$
\\ \hline
$4$&$\phi_{00}^{\otimes 3}\otimes\phi_{0n}$&$\phi_{10}^{\otimes 3}\otimes\phi_{1n}$&$\phi_{20}^{\otimes 3}\otimes\phi_{2n}$&$3\log3$
\\ \hline
\end{tabular}
\caption{Under the operation $\mathcal{B}$, what happens to $k$ copies of MES in $3\times 3$. The last column show the entanglement of $\rho^{(k)}_{3}$.}
\end{table}

\bse
\bea
&\rho^{(2)}_{3}&\stackrel{\mathcal{B}}{\longrightarrow}P_{00}\sum_{n} P_{0n}+P_{10}\sum_{n}P_{2n}+P_{20}\sum_{n}P_{1n},\\
&\rho^{(3)}_{3}&\stackrel{\mathcal{B}}{\longrightarrow}\sum_{m}P_{m0}^{\otimes 2}\sum_{n}P_{0n},\\
&\rho^{(4)}_{3}&\stackrel{\mathcal{B}}{\longrightarrow}P_{00}^{\otimes 3}\sum_{n} P_{0n}+P_{10}^{\otimes 3}\sum_{n}P_{1n}+P_{20}^{\otimes 3}\sum_{n} P_{2n}
\eea
\ese

It is required to show $\sum_{m}P_{m0}^{\otimes 2}$ is quasi-pure. $\phi_{m0}$ can be transformed to $\phi_{0m}$ in the dual basis. Performing $\mathcal{B}$ again,
$\sum_{m}P_{0m}^{\otimes 2}\stackrel{\mathcal{B}}{\longrightarrow}P_{00}\sum_{m}P_{0m}$ that is quasi-pure and $E(\rho^{(2)}_{3})=\log3$. For larger $k$, the cases above occur similarly.

{\bf Ex3} $d=4$
\begin{table}[htbp]
\begin{tabular}{|c|c|c|c|c|c|}\hline
$k$&$\phi_{0n}^{\otimes k}$&$\phi_{1n}^{\otimes k}$&$\phi_{2n}^{\otimes k}$&$\phi_{3n}^{\otimes k}$&$E(\rho_{d}^{k})$
\\ \hline\hline
$1$&$\phi_{0n}$&$\phi_{1n}$&$\phi_{2n}$&$\phi_{3n}$&$0$
\\ \hline
$2$&$\phi_{00}\otimes\phi_{0n}$&$\phi_{10}\otimes\phi_{2n}$&$\phi_{20}\otimes\phi_{0n}$&$\phi_{30}\otimes\phi_{2n}$&$\log2$
\\ \hline
$3$&$\phi_{00}^{\otimes 2}\otimes\phi_{0n}$&$\phi_{10}^{\otimes 2}\otimes\phi_{3n}$&$\phi_{20}^{\otimes 2}\otimes\phi_{2n}$&$\phi_{30}^{\otimes 2}\otimes\phi_{1n}$&$2\log4$
\\ \hline
$4$&$\phi_{00}^{\otimes 3}\otimes\phi_{0n}$&$\phi_{10}^{\otimes 3}\otimes\phi_{0n}$&$\phi_{20}^{\otimes 3}\otimes\phi_{0n}$&$\phi_{30}^{\otimes 3}\otimes\phi_{0n}$&$2\log4$
\\ \hline
$5$&$\phi_{00}^{\otimes 4}\otimes\phi_{0n}$&$\phi_{10}^{\otimes 4}\otimes\phi_{1n}$&$\phi_{20}^{\otimes 4}\otimes\phi_{2n}$&$\phi_{30}^{\otimes 4}\otimes\phi_{3n}$&$4\log4$
\\ \hline
\end{tabular}
\caption{Under the operation $\mathcal{B}$, what happens to $k$ copies of MES in $4\times 4$. The last column show the entanglement of $\rho^{(k)}_{4}$.}
\end{table}

\bse
\bea
&\rho^{(2)}_{4}&\stackrel{\mathcal{B}}{\longrightarrow}(P_{00}+P_{20})\sum_{n} P_{0n}+(P_{10}+P_{30})\sum_{n}P_{2n},\\
&\rho^{(3)}_{4}&\stackrel{\mathcal{B}}{\longrightarrow}
P_{00}^{\otimes 2}\sum_{n}P_{0n}+P_{10}^{\otimes 2}\sum_{n}P_{3n}+P_{20}^{\otimes 2}\sum_{n}P_{2n}+P_{30}^{\otimes 2}\sum_{n}P_{1n},\\
&\rho^{(4)}_{4}&\stackrel{\mathcal{B}}{\longrightarrow}\sum_{m}P_{m0}^{\otimes 3}\sum_{n} P_{0n},\\
&\rho^{(5)}_{4}&\stackrel{\mathcal{B}}{\longrightarrow}
P_{00}^{\otimes 4}\sum_{n}P_{0n}+P_{10}^{\otimes 4}\sum_{n}P_{1n}+P_{20}^{\otimes 3}\sum_{n}P_{2n}+P_{30}^{\otimes 4}\sum_{n}P_{3n}
\eea
\ese

$\rho^{(3)}_{4}$ and $\rho^{(5)}_{4}$ are exlicitly quasi-pure and $E(\rho^{(3)}_{4})=4$, $E(\rho^{(5)}_{4})=8$. It is required to explain for $\rho^{(2)}_{4}$ and $\rho^{(4)}_{4}$.  
As $\phi_{00}=(|00\rangle+|11\rangle)(|00\rangle+|11\rangle)$ and $\phi_{20}=(|01\rangle+|10\rangle)(|00\rangle+|11\rangle)$,
so $P_{00}+P_{20}=(P(|00\rangle+|11\rangle)+P(|01\rangle+|10\rangle))P(|00\rangle+|11\rangle)$ that is quasi-pure.
$P_{10}+P_{30}$ can be transformed to $P_{00}+P_{20}$ by local permutation the basis.
So $\rho^{(2)}_{4}$ is quasi-pure and $E(\rho^{(2)}_{4})=1$. 
In the dual basis, $\sum_{m}P_{m0}^{\otimes 3}$ is represented as $\sum_{n}P_{0n}^{\otimes 3}$. Further applying the operation $\mathcal{B}$, $\sum_{n}P_{0n}^{\otimes 3}\stackrel{\mathcal{B}}{\longrightarrow}P_{00}^{\otimes 2}\sum_{n}P_{0n}$ that is quasi-pure. For larger $k$, the cases above occur similarly.

{\bf Ex4} d=5
\begin{table}[htbp]
\begin{tabular}{|c|c|c|c|c|c|c|}\hline
$k$&$\phi_{0n}^{\otimes k}$&$\phi_{1n}^{\otimes k}$&$\phi_{2n}^{\otimes k}$&$\phi_{3n}^{\otimes k}$&$\phi_{4n}^{\otimes k}$&$E(\rho_{d}^{k})$
\\ \hline\hline
$1$&$\phi_{0n}$&$\phi_{1n}$&$\phi_{2n}$&$\phi_{3n}$&$\phi_{4n}$&$0$
\\ \hline
$2$&$\phi_{00}\otimes\phi_{0n}$&$\phi_{10}\otimes\phi_{2n}$&$\phi_{20}\otimes\phi_{4n}$&$\phi_{30}\otimes\phi_{1n}$&$\phi_{40}\otimes\phi_{3n}$&$\log5$
\\ \hline
$3$&$\phi_{00}^{\otimes 2}\otimes\phi_{0n}$&$\phi_{10}^{\otimes 2}\otimes\phi_{3n}$&$\phi_{20}^{\otimes 2}\otimes\phi_{1n}$&$\phi_{30}^{\otimes 2}\otimes\phi_{4n}$&$\phi_{40}^{\otimes 2}\otimes\phi_{2n}$&$2\log5$
\\ \hline
$4$&$\phi_{00}^{\otimes 3}\otimes\phi_{0n}$&$\phi_{10}^{\otimes 3}\otimes\phi_{4n}$&$\phi_{20}^{\otimes 3}\otimes\phi_{3n}$&$\phi_{30}^{\otimes 3}\otimes\phi_{2n}$&$\phi_{40}^{\otimes 3}\otimes\phi_{1n}$&$3\log5$
\\ \hline
$5$&$\phi_{00}^{\otimes 4}\otimes\phi_{0n}$&$\phi_{10}^{\otimes 4}\otimes\phi_{0n}$&$\phi_{20}^{\otimes 4}\otimes\phi_{0n}$&$\phi_{30}^{\otimes 4}\otimes\phi_{0n}$&$\phi_{40}^{\otimes 4}\otimes\phi_{0n}$&$3\log5$
\\ \hline
$6$&$\phi_{00}^{\otimes 5}\otimes\phi_{0n}$&$\phi_{10}^{\otimes 5}\otimes\phi_{1n}$&$\phi_{20}^{\otimes 5}\otimes\phi_{2n}$&$\phi_{30}^{\otimes 5}\otimes\phi_{3n}$&$\phi_{40}^{\otimes 5}\otimes\phi_{4n}$&$5\log5$
\\ \hline
\end{tabular}
\caption{Under the operation $\mathcal{B}$, what happens to $k$ copies of MES in $5\times 5$. The last column show the entanglement of $\rho^{(k)}_{5}$.}
\end{table}

\bse
\bea
&\rho^{(2)}_{5}&\stackrel{\mathcal{B}}{\longrightarrow}P_{00}\sum_{n} P_{0n}+P_{10}\sum_{n}P_{2n}+P_{20}\sum_{n}P_{4n}+P_{30}\sum_{n}P_{1n}+P_{40}\sum_{n}P_{3n},\\
&\rho^{(3)}_{5}&\stackrel{\mathcal{B}}{\longrightarrow}
P_{00}^{\otimes 2}\sum_{n}P_{0n}+P_{10}^{\otimes 2}\sum_{n}P_{3n}+P_{20}^{\otimes 2}\sum_{n}P_{1n}+P_{30}^{\otimes 2}\sum_{n}P_{4n}+P_{40}^{\otimes 2}\sum_{n}P_{2n},\\
&\rho^{(4)}_{5}&\stackrel{\mathcal{B}}{\longrightarrow}
P_{00}^{\otimes 3}\sum_{n}P_{0n}+P_{10}^{\otimes 3}\sum_{n}P_{4n}+P_{20}^{\otimes 3}\sum_{n}P_{3n}+P_{30}^{\otimes 3}\sum_{n}P_{2n}+P_{40}^{\otimes 3}\sum_{n}P_{1n},\\
&\rho^{(5)}_{5}&\stackrel{\mathcal{B}}{\longrightarrow}\sum_{m}P_{m0}^{\otimes 4}\sum_{n}P_{0n},\\
&\rho^{(6)}_{5}&\stackrel{\mathcal{B}}{\longrightarrow}\sum_{m}P_{m0}^{\otimes 5}\{\sum_{n}P_{mn}\}.
\eea
\ese

It is obvious that $\rho^{(k)}_{5}, k=2,3,4,6$ are quasi-pure and $E(\rho^{(k)}_{5})=(k-1)\log5$. 
In the dual basis, $\sum_{m}P_{m0}^{\otimes 4}$ is represented as $\sum_{n}P_{0n}^{\otimes 4}$. Further applying the operation $\mathcal{B}$, $\sum_{n}P_{0n}^{\otimes 4}\stackrel{\mathcal{B}}{\longrightarrow}P_{00}^{\otimes 3}\sum_{n}P_{0n}$.
So $\rho^{(5)}_{5}$ is quasi-pure and $E(\rho^{(5)}_{5})=3\log5$. For larger $k$, the cases above occur similarly.

{\bf Ex5} d=6
\begin{table}[htbp]
\begin{tabular}{|c|c|c|c|c|c|c|c|}\hline
$k$&$\phi_{0n}^{\otimes k}$&$\phi_{1n}^{\otimes k}$&$\phi_{2n}^{\otimes k}$&$\phi_{3n}^{\otimes k}$&$\phi_{4n}^{\otimes k}$&$\phi_{5n}^{\otimes k}$&$E(\rho_{d}^{k})$
\\ \hline\hline
$1$&$\phi_{0n}$&$\phi_{1n}$&$\phi_{2n}$&$\phi_{3n}$&$\phi_{4n}$&$\phi_{5n}$&$0$
\\ \hline
$2$&$\phi_{00}\otimes\phi_{0n}$&$\phi_{10}\otimes\phi_{2n}$&$\phi_{20}\otimes\phi_{4n}$&$\phi_{30}\otimes\phi_{0n}$&$\phi_{40}\otimes\phi_{2n}$&$\phi_{50}\otimes\phi_{4n}$&$\log3$
\\ \hline
$3$&$\phi_{00}^{\otimes 2}\otimes\phi_{0n}$&$\phi_{10}^{\otimes 2}\otimes\phi_{3n}$&$\phi_{20}^{\otimes 2}\otimes\phi_{0n}$&$\phi_{30}^{\otimes 2}\otimes\phi_{3n}$&$\phi_{40}^{\otimes 2}\otimes\phi_{0n}$&$\phi_{50}^{\otimes 2}\otimes\phi_{3n}$&$\log3+2\log2$
\\ \hline
$4$&$\phi_{00}^{\otimes 3}\otimes\phi_{0n}$&$\phi_{10}^{\otimes 3}\otimes\phi_{4n}$&$\phi_{20}^{\otimes 3}\otimes\phi_{2n}$&$\phi_{30}^{\otimes 3}\otimes\phi_{0n}$&$\phi_{40}^{\otimes 3}\otimes\phi_{4n}$&$\phi_{50}^{\otimes 3}\otimes\phi_{2n}$&$3\log3+2\log2$
\\ \hline
$5$&$\phi_{00}^{\otimes 4}\otimes\phi_{0n}$&$\phi_{10}^{\otimes 4}\otimes\phi_{5n}$&$\phi_{20}^{\otimes 4}\otimes\phi_{4n}$&$\phi_{30}^{\otimes 4}\otimes\phi_{3n}$&$\phi_{40}^{\otimes 4}\otimes\phi_{2n}$&$\phi_{50}^{\otimes 4}\otimes\phi_{1n}$&$4\log3+4\log2$
\\ \hline
$6$&$\phi_{00}^{\otimes 5}\otimes\phi_{0n}$&$\phi_{10}^{\otimes 5}\otimes\phi_{0n}$&$\phi_{20}^{\otimes 4}\otimes\phi_{0n}$&$\phi_{30}^{\otimes 5}\otimes\phi_{0n}$&$\phi_{40}^{\otimes 5}\otimes\phi_{0n}$&$\phi_{50}^{\otimes 5}\otimes\phi_{0n}$&$4\log3+4\log2$
\\ \hline
$7$&$\phi_{00}^{\otimes 6}\otimes\phi_{0n}$&$\phi_{10}^{\otimes 6}\otimes\phi_{1n}$&$\phi_{20}^{\otimes 6}\otimes\phi_{2n}$&$\phi_{30}^{\otimes 6}\otimes\phi_{3n}$&$\phi_{40}^{\otimes 6}\otimes\phi_{4n}$&$\phi_{50}^{\otimes 6}\otimes\phi_{5n}$&$6\log3+6\log2$
\\ \hline
\end{tabular}
\caption{Under the operation $\mathcal{B}$, what happens to $k$ copies of MES in $6\times 6$. The last column show the entanglement of $\rho^{(k)}_{6}$.}
\end{table}

\bse
\bea
&\rho^{(2)}_{6}&\stackrel{\mathcal{B}}{\longrightarrow}
(P_{00}+P_{30})\sum_{n}P_{0n}+(P_{10}+P_{40})\sum_{n}P_{2n}
+(P_{20}+P_{50})\sum_{n}P_{4n},\\
&\rho^{(3)}_{6}&\stackrel{\mathcal{B}}{\longrightarrow}
(P_{00}^{\otimes 2}+P_{20}^{\otimes 2}+P_{40}^{\otimes 2})\sum_{n}P_{0n}+(P_{10}^{\otimes 2}+P_{30}^{\otimes 2}+P_{50}^{\otimes 2})\sum_{n}P_{3n},\\
&\rho^{(4)}_{6}&\stackrel{\mathcal{B}}{\longrightarrow}
(P_{00}^{\otimes 3}+P_{30}^{\otimes 3})\sum_{n}P_{0n}+(P_{10}^{\otimes 3}+P_{40}^{\otimes 3})\sum_{n}P_{4n}+(P_{30}^{\otimes 3}+P_{50}^{\otimes 3})\sum_{n}P_{2n},\\
&\rho^{(5)}_{6}&\stackrel{\mathcal{B}}{\longrightarrow}
P_{00}^{\otimes 4}\sum_{n}P_{0n}+P_{10}^{\otimes 4}\sum_{n}P_{5n}+P_{20}^{\otimes 4}\sum_{n}P_{4n}+P_{30}^{\otimes 4}\sum_{n}P_{3n}+P_{40}^{\otimes 4}\sum_{n}P_{2n}+P_{50}^{\otimes 4}\sum_{n}P_{1n},\\
&\rho^{(6)}_{6}&\stackrel{\mathcal{B}}{\longrightarrow}
\sum_{m}P_{m0}^{\otimes 5}\sum_{n}P_{0n},\\
&\rho^{(7)}_{6}&\stackrel{\mathcal{B}}{\longrightarrow}\sum_{m}P_{m0}^{\otimes 6}\{\sum_{n}P_{mn}\}.
\eea 
\ese

It is apparent that $\rho^{(5)}_{6}$ and $\rho^{(7)}_{6}$ are quasi-pure. Explanation is required for $\rho^{(2)}_{6}$, $\rho^{(3)}_{6}$, $\rho^{(4)}_{6}$, and $\rho^{(6)}_{6}$. Now we show that $P_{00}+P_{30}$ is quasi-pure. 
\bea
\phi_{00}&=&(|00\rangle+|11\rangle)(|00\rangle+|11\rangle+|22\rangle),\nonumber\\
\phi_{30}&=&(|01\rangle+|10\rangle)(|00\rangle+|11\rangle+|22\rangle),\nonumber\\
P_{00}+P_{30}&=&(P(|00\rangle+|11\rangle)+P(|01\rangle+|10\rangle))P(|00\rangle+|11\rangle+|22\rangle),
\eea 
which is explicitly quasi-pure. $P_{10}+P_{40}$ and $P_{20}+P_{50}$ can be proved similarly.
So $\rho^{(2)}_{6}$ is quasi-pure and $E(\rho^{(2)}_{6})=\log3$. 
Now we show that $P_{00}^{\otimes 2}+P_{20}^{\otimes 2}+P_{40}^{\otimes 2}$ is quasi-pure. 
\bea
\phi_{00}&=&(|00\rangle+|11\rangle+|22\rangle)(|00\rangle+|11\rangle),\nonumber\\
\phi_{20}&=&(|01\rangle+|12\rangle+|20\rangle)(|00\rangle+|11\rangle),\nonumber\\
\phi_{40}&=&(|02\rangle+|10\rangle+|21\rangle)(|00\rangle+|11\rangle),\nonumber\\
P_{00}^{\otimes 2}+P_{20}^{\otimes 2}+P_{40}^{\otimes 2}&=&(P_{00}^{\otimes 2}(3)+P_{10}^{\otimes 2}(3)+P_{20}^{\otimes 2}(3))P_{00}^{\otimes 2}(2).
\eea 
From the $d=3$ case, we know that $\sum_{m}P_{m0}^{\otimes 2}(3)$ is quasi-pure. $P_{10}+P_{30}+P_{50}$ can be proved similarly.
So $\rho^{(3)}_{6}$ is quasi-pure and $E(\rho^{(2)}_{6})=\log3+2$.
Now we show $P_{00}^{\otimes 3}+P_{30}^{\otimes 3}$ is quasi-pure.
\be
P_{00}^{\otimes 3}+P_{30}^{\otimes 3}=(P_{00}^{\otimes 3}(2)+P_{10}^{\otimes 3}(2))P_{00}^{\otimes 3}(3) .
\ee
From the $d=2$ case, we know that $\sum_{m}P_{m0}^{\otimes 3}(2)$ is quasi-pure. The other two terms can be proved simliarly. So $\rho^{(4)}_{6}$ is quasi-pure and $E(\rho^{(4)}_{6})=3\log3+2$.
In the dual basis, $\sum_{m}P_{m0}^{\otimes 5}$ is represented as $\sum_{n}P_{0n}^{\otimes 5}$. Further applying the operation $\mathcal{B}$
\be
\sum_{n}P_{0n}^{\otimes 5}\stackrel{\mathcal{B}}{\longrightarrow}P_{00}^{\otimes 4}\sum_{n}P_{0n} .
\ee
So $\rho^{(6)}_{6}$ is quasi-pure and $E(\rho^{(6)}_{6})=4\log6$. For larger $k$, the cases above occur similarly. 

\begin{figure}[ht]
\begin{center}
\scalebox{0.8}{\includegraphics{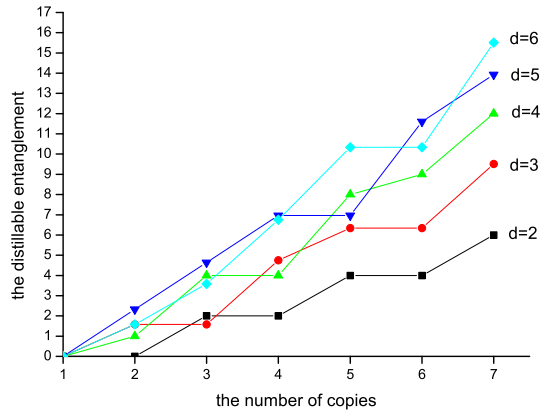}}
\end{center}
\caption{The distillable entanglement of the mixed state of $\rho_{d}^{(k)}$ is depicted in the figure. It can be seen that the distillable entanglement is monotonic in $k$ but not in $d$. A surprising fact is that $E(\rho_{d}^{(kd)})=E(\rho_{d}^{(kd-1)})$ that the terraces appear.}
\end{figure}

Notice that different conclusion is obtained \cite{Smolin} for $k=2$, where $\rho^{(2)}_{d}$ is demonstrated to be separable. And the relative entropy of entanglement is calculated based on this incorrect conclusion in a few papers. However, We prove that $\rho^{(2)}_{d}$ is distillable when $d>2$.

As well-known, irreversibility is generic in the entanglement distillation \cite{Vidal}. In this paper, we show that the mixture of multiple copies of maximally entangled states is quasi-pure and prove that the entanglement distillation is a reversible process for this nontrivial class at first glance. Whether only in the quasi-pure states is distillation reversible deserves to be considered. Another important question is how to decide the quasi-pure mixed states.\\

D. Yang thanks S. J. Gu for helpful discussion. The work is supported by the NNSF of China, the NSF of Zhejiang Province (Grant No. 602018) and Guang-Biao Cao Foundation in Zhejiang University.


\end{document}